\documentclass[article,aps,nofootinbib,twocolumn,superscriptaddress]{revtex4-1}
\input epsf
\usepackage{graphics}
\usepackage{amsmath}
\usepackage{amssymb}
\usepackage{bm}
\usepackage{color}
\usepackage{dcolumn}
\usepackage{hyphenat}
\usepackage{url}

\allowdisplaybreaks

\def\be{\begin{equation}}
\def\ee{\end{equation}}
\def\ba{\begin{eqnarray}}
\def\ea{\end{eqnarray}}

\def\ie{{\frenchspacing\it i.e.}}
\def\eg{{\frenchspacing\it e.g.}}

\newcommand{\CITE}[1]{[{\bf CITE}]}

\usepackage{graphicx}

\begin{document}

\title{Large-scale structure phenomenology of viable Horndeski theories}

\author{Simone Peirone} \affiliation{Institute Lorentz, Leiden University, PO Box 9506, Leiden 2300 RA, The Netherlands}
\author{Kazuya Koyama}\affiliation{Institute of Cosmology and Gravitation, University of Portsmouth, Portsmouth, PO1 3FX, UK}
\author{Levon Pogosian} \affiliation{Department of Physics, Simon Fraser University, Burnaby, BC, V5A 1S6, Canada}
\affiliation{Institute of Cosmology and Gravitation, University of Portsmouth, Portsmouth, PO1 3FX, UK}
\author{Marco Raveri} \affiliation{Kavli Institute for Cosmological Physics, Enrico Fermi Institute,The University of Chicago, Chicago, Illinois 60637, USA}
\author{Alessandra Silvestri} \affiliation{Institute Lorentz, Leiden University, PO Box 9506, Leiden 2300 RA, The Netherlands}

\begin{abstract}
Phenomenological functions $\Sigma$ and $\mu$, also known as $G_{\rm light}/G$ and $G_{\rm matter}/G$, are commonly used to parameterize modifications of the growth of large-scale structure in alternative theories of gravity. We study the values these functions can take in Horndeski theories, \ie~the class of scalar-tensor theories with second order equations of motion. We restrict our attention to models that are in a broad agreement with tests of gravity and the observed cosmic expansion history. In particular, we require the speed of gravity to be equal to the speed of light today, as required by the recent detection of gravitational waves and electromagnetic emission from a binary neutron star merger. We examine the correlations between the values of  $\Sigma$ and $\mu$ analytically within the quasi-static approximation, and numerically, by sampling the space of allowed solutions. We confirm that the conjecture made in~\cite{Pogosian:2016pwr}, that $(\Sigma-1)(\mu -1) \ge 0$ in viable Horndeski theories, holds very well. Along with that, we check the validity of the quasi-static approximation within different corners of Horndeski theory. Our results show that, even with the tight bound on the present day speed of gravitational waves, there is room within Horndeski theories for non-trivial signatures of modified gravity at the level of linear perturbations.
 \end{abstract}

\maketitle

\section{Introduction}
\label{sec:introduction}

A common approach to testing gravity on cosmological scales is to constrain modifications of the Einstein's equations relating the matter density contrast to the lensing and the Newtonian potentials \cite{Amendola:2007rr,Hu:2007pj,Jain:2007yk,Bertschinger:2008zb,Daniel:2008et,Zhao:2008bn,Pogosian:2010tj,Zhao:2010dz,Hojjati:2011ix,Simpson:2012ra,Silvestri:2013ne,Ade:2015rim}. The modifications, quantified in terms of functions $\Sigma$ and $\mu$, or $G_{\rm light}/G$ and $G_{\rm matter}/G$, will be well constrained by future surveys of large scale structure \cite{Zhao:2009fn,Hojjati:2011xd,Asaba:2013mxj}, such as Euclid \cite{euclid} and LSST \cite{lsst}. Given these prospects, it is pertinent to ask if measuring certain values of these functions could rule out broad classes of modified gravity (MG) theories. Moreover, in any specific MG theory, $\Sigma$ and $\mu$ would depend on the parameters of the same Lagrangian and, thus, will not be independent of each other. But are there correlations between them that hold within broad classes of theories, beyond the confines of a specific Lagrangian? We ask this question in the context of the Horndeski theories~\cite{Horndeski:1974wa,Deffayet:2011gz,Kobayashi:2011nu}, namely, all single field scalar-tensor theories with second order equations of motion.

In~\cite{Pogosian:2016pwr}, it was argued that one should expect to have $(\Sigma-1)(\mu -1) \ge 0$ in Horndeski theories that are in agreement with the existing observational and experimental constraints. In principle, mathematically, there is sufficient freedom within the Horndeski class to construct theories that would violate the conjecture. However, according to~\cite{Pogosian:2016pwr}, it would require a specially fine-tuned arrangement of separate sectors of the theory. In this paper, we set to test the conjecture made in~\cite{Pogosian:2016pwr} by numerically sampling the space of viable Horndeski models. In addition, we would like to better understand properties of the models that happen to violate the conjecture.

To sample the space of solutions of Horndeski theories, we use the so-called EFT approach~\cite{Gubitosi:2012hu,Bloomfield:2012ff,Gleyzes:2013ooa,Bloomfield:2013efa,Piazza:2013coa} to modeling scalar field dark energy. In the EFT approach, solving for the background evolution and linear perturbations in Horndeski theories requires specifying five functions of time. Two of these functions affect both the background and the perturbations, while the other three concern only the perturbations. An ensemble of viable Horndeski models can be obtained by randomly generating the five EFT functions and keeping those that lead to theoretically consistent and observationally allowed solutions.

A similar numerical test was performed in~\cite{Perenon:2016blf}, which, however, was based on an alternative way of formulating the EFT~\cite{Bellini:2014fua,Gleyzes:2014rba}. There, the expansion history was provided independently from the four functions that determine the evolution of linear perturbations. This amounts to the assumption that the modification of the evolution of perturbations is uncorrelated with the changes to the background expansion. However, in any theory, the expansion history and the perturbations are derived from the same Lagrangian and thus they must be partially correlated. In our approach, where two of the five independent functions control both the background and perturbations, requiring the expansion history to be in a broad agreement with observations makes it more challenging to fine-tune an arrangement where $(\Sigma-1)(\mu -1) < 0$. 

The detection of gravitational waves (GW170817) and the associated gamma-ray bursts (GRB170817A) from a neutron star merger \cite{TheLIGOScientific:2017qsa,Monitor:2017mdv,Coulter:2017wya} has put stringent constraints on the difference between the speed of light and gravitational waves. This has a significant implication for modified gravity models, in particular scalar tensor theories~\cite{Lombriser:2015sxa,Lombriser:2016yzn,Brax:2015dma, Bettoni:2016mij,Creminelli:2017sry,Sakstein:2017xjx,Ezquiaga:2017ekz,Baker:2017hug, Arai:2017hxj, Gong:2017kim, Jana:2017ost, Amendola:2017orw, Crisostomi:2017lbg,Langlois:2017dyl,Heisenberg:2017qka,Hou:2017cjy}.
As we will show in this paper, there is still ample room for modified gravity models to predict $\Sigma-1 \neq 0$ and $\mu- 1 \neq 0$ on large scale structure scales. Requiring the present value of the speed of gravitational waves to be equal to the speed of light further restricts the space of opportunities for violating the $(\Sigma-1)(\mu -1) \ge 0$ conjecture.

The conjecture in~\cite{Pogosian:2016pwr} was based on explicit expressions for $\Sigma$ and $\mu$ derived under the Quasi Static Approximation (QSA). Since our numerical procedure allows us to compute these functions exactly, we verify the validity of the QSA at several representative scales and redshifts. We find that the QSA breaks down at $k \lesssim 0.001$ h/Mpc even though the modes are still well-within the scalar field sound horizon, indicating that the time derivatives of the metric and the scalar field perturbations can no longer be neglected on those scales. Nevertheless, we find that the $(\Sigma-1)(\mu -1) \ge 0$ conjecture holds very well on scales probed by large scale structure surveys.

Our work demonstrates the complementarity of the purely phenomenological $\Sigma$ and $\mu$ parameterization and the EFT approach to testing scalar-tensor theories. The latter can be used to derive theoretical priors on $\Sigma$ and $\mu$, which are more directly constrained by observations.

In what follows, we review the phenomenological description of cosmological perturbations in Horndeski theories in Section~\ref{sec:definitions} and  analytically examine the conditions for violating $(\Sigma-1)(\mu -1) \ge 0$ in Section~\ref{sec:conjecture}. We describe the procedure and present the results of the numerical sampling of $\Sigma$ and $\mu$ in three representative subclasses of Horndeski theories in Section~\ref{sec:ensembles} and conclude with a discussion in Section~\ref{sec:summary}.

\section{$\Sigma$ and $\mu$ in Horndeski theories}
\label{sec:definitions}

In the Newtonian gauge, scalar perturbations to the Friedmann-Lemaitre-Robertson-Walker (FLRW) metric are the gravitational potentials $\Psi$ and $\Phi$, defined via
\be
ds^2 = -(1+2\Psi)dt^2 + a^2(1-2\Phi) d{\bf x}^2 \ ,
\ee
where $a$ is the scale factor. As discussed in \cite{Zhang:2007nk,Amendola:2007rr}, non-relativistic particles respond to gradients of $\Psi$, while relativistic particles ``feel'' the gradients of the Weyl potential, $(\Phi +\Psi)/2$. In LCDM, at epochs when the radiation density can be neglected, one has $(\Phi +\Psi)/2= \Phi = \Psi$. However, in alternative models, in which additional degrees of freedom can mediate gravitational interactions, the potentials need not be equal. It will be possible to test this by combining the weak lensing shear and galaxy redshift data from surveys like Euclid \cite{euclid} and LSST \cite{lsst}. A common practical way of conducting such tests \cite{Pogosian:2010tj} involves introducing phenomenological functions $\mu$ and $\Sigma$, defined as
\ba
 k^2 \Psi &=& -4\pi G \mu (a,k) a^2 \rho \Delta \ ,
\label{poisson_mg} \\
 k^2(\Phi +\Psi) &=& -8\pi G\, \Sigma(a,k)\, a^2 \rho \Delta \ ,
\label{weyl_mg}
\ea

\noindent
where $\rho$ is the background matter density and $\Delta$ is the comoving density contrast~\footnote{$\Delta = \delta +3 a H v/k$, where $\delta$ is the overdensity in the Newtonian conformal gauge, $v$ is the irrotational component of the peculiar velocity and $H$ is the Hubble function.}.
Alternatively, one could use any one of the above functions along with the ``gravitational slip'' \cite{Amendola:2007rr,Hu:2007pj,Jain:2007yk,Bertschinger:2008zb,Daniel:2008et} $\gamma(a,k)$ defined via $\Phi = \gamma(a,k) \Psi$. As shown in \cite{Hojjati:2011xd,Asaba:2013mxj}, $\Sigma$ will be well-constrained by the combination of weak lensing and photometric galaxy counts from surveys like Euclid and LSST. Spectroscopic galaxy redshifts will add measurements of redshift space distortions, which probe the Newtonian potential, and will help to measure $\mu$ \cite{Song:2010fg,Simpson:2012ra,Asaba:2013mxj}. The parameter $\gamma$ is not directly probed by cosmological observables but can be derived from the measurement of the other two. 

Given a parameterization of $\Sigma$ and $\mu$, one can solve for the evolution of cosmological perturbations~\cite{Pogosian:2010tj} using, \eg, the publicly available code MGCAMB \cite{Zhao:2008bn,Hojjati:2011ix}, and constrain the parameters by fitting them to data. The question one should then ask is if the measured values of the parameters rule out certain classes of modified gravity models.

Obtaining a closed functional form of $\Sigma$ and $\mu$ in a given gravity theory is only possible under the quasi-static (QS) approximation (QSA). The QSA has been shown to hold well in certain representative classes of scalar-tensor theories \cite{Deffayet:2009wt,Brax:2012gr,Hojjati:2012rf,Barreira:2012kk,Noller:2013wca}.

In \cite{Pogosian:2016pwr}, the QS expressions for $\Sigma$ and $\mu$ in the Horndeski class of scalar-tensor theories were derived and closely examined. It was observed that there must be correlations between their values. In particular, one should generally expect to have $\Sigma-1$ and $\mu-1$ to be of the same sign in theoretically consistent models that do not grossly contradict observations. We revisit this conjecture in Section \ref{sec:conjecture} after briefly reviewing the EFT description of the Horndeski theories and the QS forms of $\Sigma$ and $\mu$ in the remainder of this section.

\subsection{Horndeski theories and their EFT description}

The action of the most general scalar-tensor theory with second order equations of motion, also known as the Hordneski class of theories \cite{Horndeski:1974wa,Deffayet:2011gz,Kobayashi:2011nu}, can be written as
\begin{equation}
S = \int d^4x \sqrt{-g} \left[\sum_{i=2}^{5}{\cal L}_{i} + {\cal L}_M(g_{\mu \nu},\chi_m) \right] \ ,
\label{S_Horndeski}
\end{equation}
with
\begin{eqnarray}
\nonumber
{\cal L}_{2} & = & K(\phi,X),\\ \nonumber
{\cal L}_{3} & = & -G_{3}(\phi,X)\Box\phi,\\ \nonumber
{\cal L}_{4} & = & G_{4}(\phi,X)\, R+G_{4X}\,[(\Box\phi)^{2}-(\nabla_{\mu}\nabla_{\nu}\phi)\,(\nabla^{\mu}\nabla^{\nu}\phi)]\,,\\ \nonumber
{\cal L}_{5} & = & G_{5}(\phi,X)\, G_{\mu\nu}\,(\nabla^{\mu}\nabla^{\nu}\phi) \\ \nonumber
&-&\frac{1}{6}\, G_{5X}\,[(\Box\phi)^{3}-3(\Box\phi)\,(\nabla_{\mu}\nabla_{\nu}\phi)\,(\nabla^{\mu}\nabla^{\nu}\phi) \\
&+&2(\nabla^{\mu}\nabla_{\alpha}\phi)\,(\nabla^{\alpha}\nabla_{\beta}\phi)\,(\nabla^{\beta}\nabla_{\mu}\phi)] \ ,
\label{Horn_Li}
\end{eqnarray}
where $K$ and $G_{i}$ ($i=3,4,5$) are functions of the scalar field $\phi$ and its kinetic energy $X=-\partial^{\mu}\phi\partial_{\mu}\phi/2$, $R$ is the Ricci scalar, $G_{\mu\nu}$ is the Einstein tensor, $G_{iX}$ and $G_{i\phi}$ denote the partial derivatives of $G_{i}$ with respect to $X$ and $\phi$, respectively, and ${\cal L}_M(g_{\mu \nu},\chi_m)$ is the Lagrangian for matter fields, collectively denoted with $\chi_m$, minimally coupled to the metric $g_{\mu \nu}$. 

A general way to model the background evolution and linear perturbations in a wide class of scalar field models was proposed in~\cite{Gubitosi:2012hu,Bloomfield:2012ff} and further developed in~\cite{Gleyzes:2013ooa,Bloomfield:2013efa,Piazza:2013coa}. For the class of Horndeski theories, the EFT action is 
\ba
\mathcal{S} =&& \int d^4x \sqrt{-g}  \bigg\{ \frac{m_0^2}{2} \Omega(t)R + \Lambda(t) - c(t)\,a^2\delta g^{00}  \nonumber \\ 
+&& \frac{M_2^4 (t)}{2} \left( a^2\delta g^{00} \right)^2
 - \frac{\bar{M}_1^3 (t)}{2} \, a^2\delta g^{00}\,\delta {K}{^\mu_\mu}  \nonumber \\
+&&  \frac{\bar{M}_2^2 (t)}{2}\left[\left( \delta {K}{^\mu_\mu}\right)^2 - \delta {K}{^\mu_\nu}\,\delta {K}{^\nu_\mu}-\frac{ a^2}{2} \delta g^{00}\,\delta \mathcal{R}\right]+	\ldots \bigg\}\nonumber\\
+&& S_{m} [g_{\mu \nu}, \chi_m ],
\label{EFT_action}
\ea
where $m_0^{-2} = 8\pi G$, and $\delta g^{00}$, $\delta {K}{^\mu_\nu}$, $\delta K$ and $\delta R^{(3)}$ are, respectively, the perturbations of the time-time component of the metric, the extrinsic curvature and its trace, and the three dimensional spatial Ricci scalar of the constant-time hypersurfaces. The action (\ref{EFT_action}) is written in the  unitary gauge, in which the time coordinate is associated with hypersurfaces of a uniform scalar field. The EFT functions $\Omega$, $\Lambda$, $c$, $\bar{M}^3_1$, $M_2^4$, $\bar{M}^2_2$ appearing in (\ref{EFT_action}) can be expressed in terms of the functions appearing in the Horndeski Lagrangian (\ref{Horn_Li}) \cite{Bloomfield:2013efa}. The first three functions, $\Omega$, $\Lambda$ and $c$, affect both the background and the perturbations, with only two of them being independent (one function can be solved for by using the two Friedmann equations). The remaining three functions, $\bar{M}^3_1$, $M_2^4$ and $\bar{M}^2_2$, concern only the perturbations. 

An equivalent alternative way of parameterizing the EFT action for linear perturbations around a given FLRW background in Horndeski models is based on the following action for linear perturbations \cite{Bellini:2014fua,Gleyzes:2014rba,Gleyzes:2015rua,Gleyzes:2015pma}:
\ba
\nonumber
 S^{(2)} &=& \int dtdx^3\, a^3 {M_*^2 \over 2} \bigg\{ \delta K^i_j \delta K^j_i - \delta K^2 + R \delta N \\ \nonumber
 &+& (1+ \alpha_T) \delta_2 \left( \sqrt{h}R/a^3 \right) + \alpha_K H^2 \delta N^2 \\
 &+& 4\alpha_B H \delta K \delta N  \bigg\}+ S^{(2)}_{m} [g_{\mu \nu}, \chi_m] \ ,
\label{alpha_action}
\ea
where $N$ is the lapse function and $S^{(2)}_{m}$ is the action for matter perturbations in the Jordan frame. This action is parameterized by five functions of time: the Hubble rate $H$, the generalized Planck mass $M_*$, the gravity wave speed excess $\alpha_T$, the ``kineticity'' $\alpha_K$, and the ``braiding'' $\alpha_B$ \cite{Bellini:2014fua}. One also defines a derived function, $\alpha_M$, which quantifies the running of the Planck mass. The relations between the functions in the two EFT approaches are provided in the Appendix. 

We emphasize a key difference between the two EFT descriptions. In the first, the expansion history is derived, given the EFT functions. In the second approach, $H(a)$ is treated as one of the independent functions that needs to be provided. This distinction is important when it comes to sampling the viable solutions of Horndeski theories, as it amounts to a different choice of priors. 

\subsection{$\Sigma$ and $\mu$ in Horndeski theories}

The theoretical expressions for $\mu$ and $\Sigma$ can be derived under the QSA, where one considers the scales below the scalar field sound horizon and ignores the time-derivatives of the scalar field perturbations and the gravitational potentials. In Horndeski theories, they have the form of a ratio of quadratic polynomials in $k$ \cite{DeFelice:2011hq,Silvestri:2013ne,Pogosian:2016pwr}:
\ba
\mu &=& {m_0^2 \over M_*^2} {1+ M^2 \ a^2/k^2 \over f_3/2f_1M_*^2+M^2(1+\alpha_T)^{-1} a^2/k^2},
\label{eq:mu-horn} \\
\Sigma &=& {m_0^2 \over 2M_*^2} {1+f_5/f_1+M^2[1+(1+\alpha_T)^{-1}] a^2/k^2 \over f_3/2f_1M_*^2+M^2(1+\alpha_T)^{-1} a^2/k^2},
\label{eq:Sigma-horn}
\ea
where we defined $M^2 \equiv C_\pi /f_1$ and with the functions $C_\pi$,  $f_1$, $f_3$ and $f_5$ defined in Appendix \ref{app:hor}. The mass parameter $M$ sets the scale below which the scalar field fluctuations contribute a fifth force, \ie, the Compton wavelength $\lambda_C \sim M^{-1}$.

\section{The $(\Sigma -1)(\mu -1) \ge 0$ conjecture}
\label{sec:conjecture}

In \cite{Pogosian:2016pwr}, it was conjectured that viable Horndeski models should have
\be
(\Sigma-1)(\mu-1) \ge 0 \ .
\label{eq:conjecture}
\ee
Mathematically, there is sufficient freedom in Horndeski theories to violate (\ref{eq:conjecture}). The conjecture is such that violations are unlikely, because they require balancing the evolution of the background gravitational coupling, \ie~the $m_0^2/M_*^2$ pre-factor in Eqs.~(\ref{eq:mu-horn})-(\ref{Sigma_inf}), with the change in the speed of gravity waves ($\alpha_T$) and the fifth force contribution, quantified by $\beta_B$ and $\beta_\xi$, in a rather special way. A statement about the likeliness of something occurring necessarily depends on the choice of the priors. In this instance, the key assumption is that the dynamics of both the background and the perturbations are derived from the same Lagrangian, which can be of any form consistent with (\ref{Horn_Li}). For instance, one could imaging constructing an ensemble of Horndeski theories by randomly sampling all functions of $\phi$ and $X$ appearing in (\ref{Horn_Li}), along with all possible initial conditions. Since an evolving gravitational coupling affects both the expansion rate and the fifth force contribution, restricting to the subset of solutions with an acceptable $H(a)$ reduces the probability of achieving the fine-tuning necessary to violate (\ref{eq:conjecture}).

In practice, sampling the action (\ref{Horn_Li}) directly would be prohibitively costly without making significant simplifying assumptions (\eg~see \cite{Arai:2017hxj}). Another option, given that we are only interested in the background and linear perturbations, is to work with (\ref{EFT_action}) and sample the EFT functions, treating them as being {\it a priori} independent. Since functions $\Omega$ and $\Lambda$ (and $c$, which can be derived from them) in (\ref{EFT_action}) affect the background evolution, {\it a posteriori} restrictions on $H(a)$ will constrain variations in $\Omega(a)$, which is the EFT function controlling the evolution of the gravitational coupling, making it harder to violate the conjecture (\ref{eq:conjecture}). This effect would be absent had we assumed that $H(a)$ was known {\it a priori}, which is the case if one samples the action (\ref{alpha_action}) instead, where $H(a)$ is assumed to be known independently from $M^2_*(a)$, $\alpha_B$, $\alpha_K$ and $\alpha_T$. The probability of seeing exceptions to (\ref{eq:conjecture}) is further lowered by constraints on the variation of the gravitational coupling from the big bang nucleosynthesis (BBN), CMB and various fifth force bounds \cite{Uzan:2010pm}, and the strict bound on the speed of gravitational waves imposed by GW170817 and GRB170817A \cite{TheLIGOScientific:2017qsa,Monitor:2017mdv,Coulter:2017wya}.

In the remainder of this section, we analytically examine the conditions under which (\ref{eq:conjecture}) can be violated, separately considering the limiting cases of the super- and sub-Compton scales. It is reasonable to expect the cosmological observational window to fall into one of these limits, since the Compton wavelength is either very large ($\lambda_C \sim H^{-1}$) in models of self-accelerating type \cite{Deffayet:2009wt}, or very small ($\lambda_C < 1$ Mpc) in models of chameleon type \cite{Capozziello:2003tk,Carroll:2003wy,Appleby:2007vb,Hu:2007nk,Starobinsky:2007hu,Khoury:2003aq,Hinterbichler:2010es,Damour:1994zq,Brax:2011ja}. The exact solutions can be studied numerically and are presented in Section \ref{sec:ensembles}.

\subsection{The super-Compton limit}

In the $k/a \ll M$ limit, corresponding to scales above the Compton wavelength, (\ref{eq:mu-horn}) and (\ref{eq:Sigma-horn}) reduce to
\ba
\mu_0 &=&{m_0^2 \over M_*^2}(1+\alpha_T), \label{mu0_Hornd} \\
\Sigma_0 &=& {m_0^2 \over M_*^2} \left(1+{\alpha_T \over 2} \right). 
\label{Sigma0_Hornd}
\ea
This implies that the gravitational slip on super-Compton scales is determined solely by the speed of gravitational waves \cite{Pogosian:2016pwr}, \ie
\be
\gamma_0 = {1 \over 1+\alpha_T} = c_T^{-2}.
\label{gamma0_Hornd}
\ee
The condition to have $\mu_0 > 1$ and $\Sigma_0 < 1$ can be written as
\be
(1+\alpha_T)(1+{1\over 2}\alpha_T) < \Omega < (1+\alpha_T)^2,
\label{m0gt1S0lt1}
\ee
where we have used Eqs.~(\ref{eq:M*}) and~(\ref{eq:alphaT}) to express $M_*^2$ in (\ref{Sigma0_Hornd}) in terms of $\Omega$ and $\alpha_T$. A necessary condition for (\ref{m0gt1S0lt1}) to hold is $\alpha_T > 0$, which implies $\Omega>1$. Similarly, to have $\mu_0 < 1$ and $\Sigma_0 > 1$, we must have
\be
(1+\alpha_T)^2 < \Omega < (1+\alpha_T)(1+ {1\over 2}\alpha_T),
\label{m0lt1S0gt1}
\ee
which requires $\alpha_T < 0$ and, hence, $\Omega<1$. The conditions (\ref{m0gt1S0lt1}) and (\ref{m0lt1S0gt1}) imply that, to have an observable violation of  (\ref{eq:conjecture}), there has to be a significant $\alpha_T \ne 0$ and a corresponding $\Omega \ne 1$, both of which are constrained to be close to their GR values today \cite{Caves:1980jn,Moore:2001bv,Jimenez:2015bwa}. While GW170817 and GRB170817A \cite{TheLIGOScientific:2017qsa,Monitor:2017mdv,Coulter:2017wya} require $\alpha_T$ to vanish at $z<0.01$,  in principle, there are no observational bounds on $\alpha_T$ at high redshifts. On the other hand, $\Omega$ is constrained to be within $10$\% of its today's value during the BBN epoch and at the last scattering \cite{Uzan:2010pm}. Also, $\dot \Omega \ne 0$ implies a new interaction between massive particles mediated by the scalar field, which is constrained by probes of structure formation. Thus, it would be challenging to arrange for (\ref{eq:conjecture}) to be violated on super-Compton scales, \emph{and} be observable.

\subsection{The sub-Compton limit}

On scales below the Compton wavelength, {\it i.e.} in the limit $k/a \gg M$, the expressions for $\mu$ and $\Sigma$ become
\ba
\mu_\infty &=&{m_0^2 \over M_*^2}(1+\alpha_T+\beta_\xi^2), \label{mu_inf} \\
\Sigma_\infty &=& {m_0^2 \over M_*^2} \left(1+{\alpha_T \over 2} + {\beta_\xi^2 + \beta_B \beta_\xi \over 2} \right). 
\label{Sigma_inf}
\ea
where, following \cite{Gleyzes:2015rua}\footnote{The definition of $\alpha_B$ in \cite{Gleyzes:2015rua} differs from that in \cite{Bellini:2014fua} by a factor of $-2$. We use the original definition of \cite{Bellini:2014fua}.}, we defined
\ba
\beta_B &=& -\sqrt{2\over c^2_s \alpha} \ {\alpha_B \over 2} \\
\beta_\xi &=& \sqrt{2 \over c^2_s \alpha} \left[-{\alpha_B \over 2}(1+\alpha_T)+\alpha_T-\alpha_M \right] \\
\alpha &=& \alpha_K+{3\over 2}\alpha_B^2 \ ,
\ea
with the expression for the speed of sound of the scalar field perturbations, $c_s^2$, given by Eq.~(\ref{eq:cs2}) in Appendix \ref{app:hor}. Stability of linear perturbations requires $\alpha>0$ and $c_s^2>0$ \cite{Hu:2014oga,Bellini:2014fua}.

The condition to have $\mu > 1$ and $\Sigma < 1$ is
\be
1+{1\over 2}(\alpha_T+\beta_\xi^2 + \beta_\xi \beta_B) < {\Omega \over 1+ \alpha_T} < 1+\alpha_T + \beta_\xi^2 \ ,
\label{eq:condition1}
\ee
while, to have $\mu < 1$ and $\Sigma > 1$, we must have
\be
1+\alpha_T + \beta_\xi^2 <  {\Omega \over 1+ \alpha_T}  < 1+{1\over 2}(\alpha_T+\beta_\xi^2 + \beta_\xi \beta_B) \ .
\label{eq:condition2}
\ee
The argument made in \cite{Pogosian:2016pwr} was that it would take significant fine-tuning to arrange for the background ($\Omega, \alpha_T$) contributions to $\mu$ and $\Sigma$ to balance the fifth force ($\beta_\xi, \beta_B$) contributions in a precise way to satisfy conditions (\ref{eq:condition1}) or (\ref{eq:condition2}).

To gain insight into the degree of fine-tuning involved in satisfying conditions (\ref{eq:condition1}) or (\ref{eq:condition2}), we next examine the subclass of theories with $\alpha_T=0$. Such theories are simpler to analyze and are favored by the recent bounds from GW170817 and GRB170817A \cite{TheLIGOScientific:2017qsa,Monitor:2017mdv,Coulter:2017wya}.

\subsection{Theories with unmodified speed of gravitational waves}

We will refer to the sub-class of Horndeski theories with the speed of gravity equal to the speed of light as $H_S$. The change in the gravity speed is given by $\alpha_T$, related to EFT functions via
\be
\alpha_T =  -\bar{M}^2_2/M_*^2 \ .
\ee
Setting $\bar{M}^2_2=0$ within the EFT framework ensures $\alpha_T=0$. In terms of the functions in the Horndeski Lagrangian, $\alpha_T$ is given by \cite{Bellini:2014fua}
\be
\alpha_T = 2X[2G_{4X}-2G_{5\phi}-(\ddot{\phi} - H\dot{\phi}) G_{5X}] M_*^{-2} \ .
\ee
Thus, requiring $\alpha_T=0$ implies $G_{4X}=G_{5X}=G_{5\phi}=0$ as discussed in~\cite{Pogosian:2016pwr} and more recently in~\cite{Creminelli:2017sry,Ezquiaga:2017ekz}.An example of models with non trivial kinetic term that satisfy such condition is the Kinetic Gravity Brading gravity~\cite{Deffayet:2010qz}.

In $H_S$, the non-trivial EFT functions are $\Omega$, $\Lambda$, $c$, $M_2^4$ and $\bar{M}_1^3$. Using the relations (\ref{eq:M*})-~(\ref{eq:alphaK}), we can write
\ba
M_*^2 &=& m_0^2\Omega \\
\alpha_M &=& {\dot{\Omega} \over H\Omega} \\
\alpha_B &=& - {\dot{\Omega} \over H\Omega}  - {\bar{M}^3_1 \over H m_0^2\Omega} = -\alpha_M - g_3,
\ea
where we have introduced 
\be
g_3 \equiv {\bar{M}^3_1 \over H m_0^2\Omega} \ .
\ee
Then,
\ba
\beta_B &=& 
\sqrt{2\over c^2_s \alpha}  {\alpha_M+g_3\over2} \\
\beta_\xi &=& 
\sqrt{2\over c^2_s \alpha}\left[{g_3 -\alpha_M \over 2} \right].
\ea
Substituting these expressions into Eqs.~(\ref{mu_inf}) and (\ref{Sigma_inf}), we get
\be
\mu_\infty = 
{1 \over \Omega} \left[ 1+ \nu (\alpha_M -g_3)^2 \right]  \ ,
\label{mu_infty_hs}
\ee
and
\ba
\nonumber
\Sigma_\infty &=& 
{1 \over \Omega} \left[ 1+ \nu (\alpha_M -g_3)^2 + \nu (\alpha_M g_3 - \alpha_M^2) \right] \\
&=& \mu_\infty + {\nu \over \Omega} (\alpha_M g_3 - \alpha_M^2)
 \ ,
 \label{eq:sigma_g3}
\ea
where we have defined $\nu \equiv (2 c_s^2 \alpha)^{-1}$. Conditions (\ref{eq:condition1}) and (\ref{eq:condition2}) become
\be
1+ \nu (g_3^2 - \alpha_M g_3 ) < \Omega < 1 + \nu (\alpha_M -g_3)^2
\label{eq:condition1g}
\ee
and
\be
1 + \nu (\alpha_M -g_3)^2 < \Omega < 1+ \nu (g_3^2 - \alpha_M g_3 ) .
\label{eq:condition2g}
\ee
In addition, stability conditions require $c_s^2 \alpha \ge 0$, hence $\nu$ cannot be negative.

At this point, we can make two observations:
\begin{enumerate}
\item Neither (\ref{eq:condition1g}) nor (\ref{eq:condition2g}) can be satisfied if $\alpha_M \propto \dot{\Omega} =0$. Thus, violating the conjecture generally requires a notable variation of the background gravitational coupling, which is observationally constrained \cite{Uzan:2010pm};
\item Condition (\ref{eq:condition2g}) cannot be satisfied if $g_3=0$, implying that $\mu < 1$ and $\Sigma>1$ cannot happen in models with a canonical form of the scalar field kinetic energy term, {\it i. e.} models of the generalized Brans-Dicke (GBD) type.
\end{enumerate}
To gain further insight, let us consider conditions (\ref{eq:condition1g}) and (\ref{eq:condition2g}) separately.

\subsubsection{Conditions for having $\mu > 1$ and $\Sigma<1$}
\label{sec:conjectureC1}

Since $\nu$ is non-negative, a necessary condition for (\ref{eq:condition1g}) to hold is $(\alpha_M - g_3)^2 > (g_3^2 - \alpha_M g_3)$, or
\be
\alpha_M^2 > \alpha_M g_3,
\ee
which is automatically satisfied if $\alpha_M$ and $g_3$ have opposite signs. In principle, there is nothing prohibiting this from happening. However, observational constraints on $\Omega$ and $\alpha_M \propto \dot{\Omega}$, as well as constraints on $H(a)$ which also limit variations of $\Omega(a)$, will generally suppress large departures from GR with $\mu > 1$ and $\Sigma<1$. This is, in fact, what we see in our simulations, comparing the results before and after the observational constraints are applied.

\subsubsection{Conditions for having $\mu < 1$ and $\Sigma>1$}
\label{sec:conjectureC2}

Requiring stability of perturbations plays an important role in eliminating solutions with $\mu < 1$ and $\Sigma>1$. Stability ensures that the force mediated by the scalar field fluctuations is attractive, thus increasing the value of the effective Newton's constant. The only way to arrange for $\mu<1$ is by making $\Omega>1$. But $\Omega$ is constrained to be close to unity today \cite{Brax:2012gr,Joyce:2014kja,Perenon:2015sla}, which means it would be very difficult to detect $\mu < 1$ at low redshifts. Having $\Omega>1$ would also tend to make $\Sigma <1$, unless the fifth force contribution to $\Sigma$ is large enough to make $\Sigma>1$, while still being small enough to keep $\mu<1$, which is hard to arrange.

Mathematically, a necessary condition for (\ref{eq:condition2g}) to hold is $(\alpha_M - g_3)^2 < (g_3^2 - \alpha_M g_3)$, or
\be
\alpha_M^2 < \alpha_M g_3 \ .
\label{condQ4}
\ee
This is satisfied only if $\alpha_M$ and $g_3$ are of the same sign and $\alpha_M^2<g_3^2$. On the other hand, stability of perturbations requires $c_s^2\alpha>0$, which, for $H_S$, can be written as
\ba
\nonumber
c_s^2\alpha = (\alpha_M^2 -g_3^2) &+&2(\alpha_M-g_3) - {2\dot{H} \over H^2} (2+\alpha_M+g_3)  \\
&-& {1\over H}(\dot{\alpha}_M+\dot{g_3}) - {\rho_m+P_m \over M^2_*H^2} >0.
\label{stabH3}
\ea
Note that $\alpha_M^2<g_3^2$ makes the first term on the right hand side of (\ref{stabH3}) strictly negative, while the other terms could still be of either sign. Now, imagine sampling $\alpha_M$ and $g_3$ from a distribution centered around $0$. The strictly negative first term would skew $c_s^2\alpha$ towards negative values, reducing the probability of simultaneously satisfying (\ref{condQ4}) and (\ref{stabH3}). In the next Section, we numerically confirm that imposing the stability condition practically eliminates the solutions with $\mu < 1$ and $\Sigma>1$.

\section{The ensemble of $\mu$ and $\Sigma$ in Horndeski theories}
\label{sec:ensembles}

\begin{table*}[t]
{\renewcommand{\arraystretch}{1.5}
\begin{tabular}{|c|c|c|c|}
\hline
Name & \ Lagrangian functions in (\ref{Horn_Li}) \  & \ ``EFT'' functions in (\ref{EFT_action}) \  & \ ``Unified'' functions in (\ref{alpha_action}) \ 
\\[1ex] \hline \hline
GBD & \ $K=X-V(\phi)$,  $G_4=G_4(\phi)$\ & \ $\Omega$, $\Lambda$  \  & \ $H$, $\alpha_B=-\alpha_M$, $\alpha_K$ \ \\ \hline 
$H_S$ & \ $K(X,\phi)$,  $G_3(X,\phi)$,  $G_4=G_4(\phi)$ \ & \ $\Omega$, $\Lambda$, $\bar{M}^3_1$, $M_2^4$   \  & \ $H$, $\alpha_B$, $\alpha_M$, $\alpha_K$ \ \\ \hline 
Horndeski & \ $K(X,\phi)$, \ $G_i(X,\phi)$, \ $i=3,4,5$ \ & \ $\Omega$, $\Lambda$, $\bar{M}^3_1$, $M_2^4$, $\bar{M}^2_2(z=0)=0$  \  & \ $H$, $\alpha_B$, $\alpha_M$, $\alpha_K$, $\alpha_T(z=0)=0$ \ \\ \hline 
 \end{tabular}}
\caption{The three sub-classes of Horndeski theories considered in Section \ref{sec:ensembles}.}
\label{tab:models}
\end{table*}

\begin{figure}[htbp]
\includegraphics[width=0.45\textwidth]{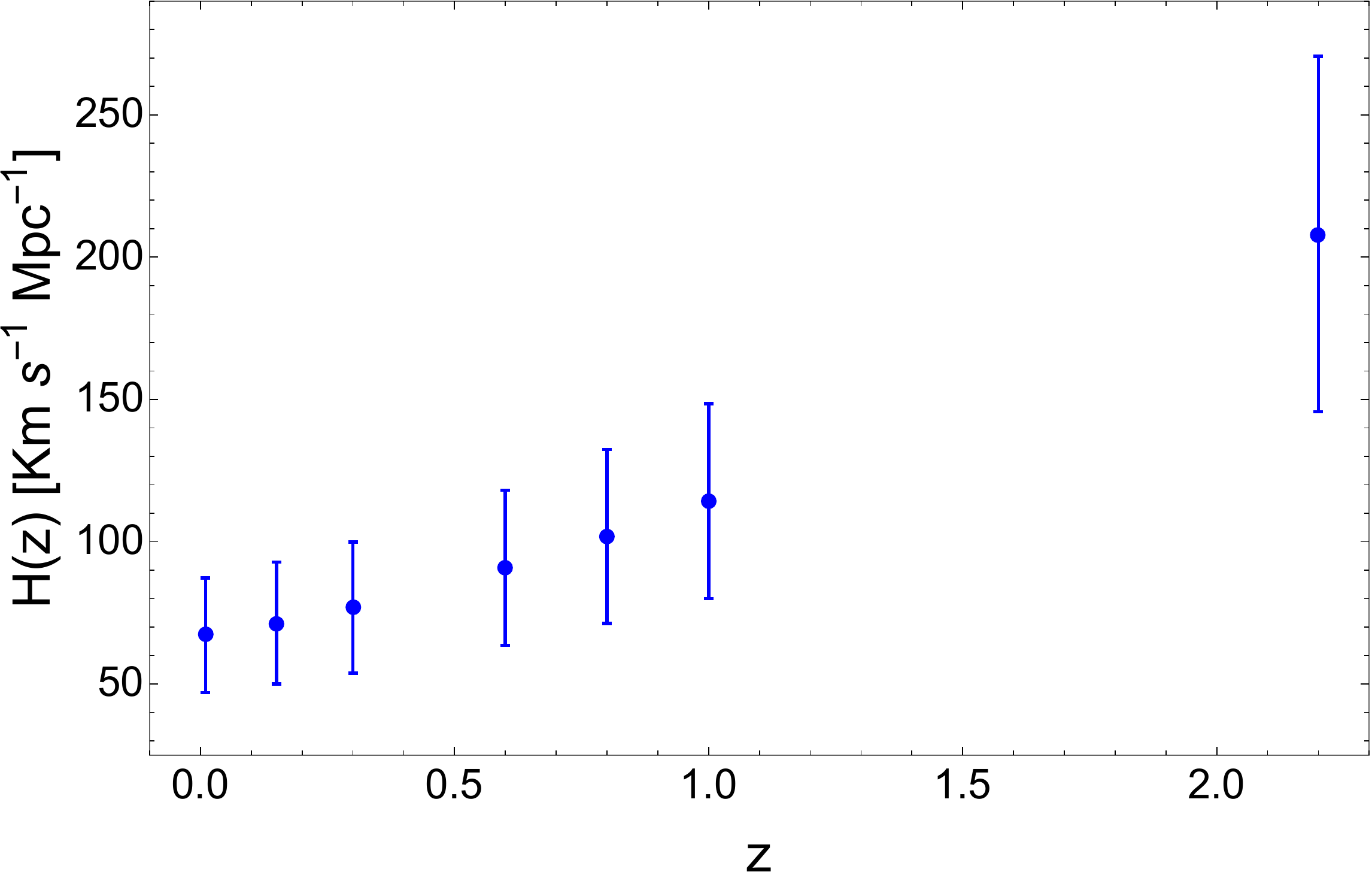}
\caption{The peak values and the standard deviation of the Gaussian prior imposed on the evolution of the Hubble parameter, $H(z)$. The fiducial expansion history corresponds to the Planck 2015 best fit $\Lambda$CDM model \cite{Ade:2015xua}. The standard deviation is chosen to be wide enough to accommodate any tensions that may exist between different datasets.}
\label{fig:H-prior}
\end{figure}

We have performed a numerical simulation to check if there are notable correlations between values of $\Sigma$ and $\mu$, and if they are consistent with the analytical arguments presented in the previous section. To this end, we have generated an ensemble of EFT functions and, for each realization, evaluated $\Sigma$ and $\mu$ at different $k$ and $a$, along with the corresponding background expansion history $H(a)$. Then we checked if $(\Sigma -1)(\mu -1) \ge 0$ holds for viable models from the ensemble. 

Following \cite{Raveri:2017qvt}, we parametrize the EFT functions using Pad\'e functions,
\begin{align}
f(a) = \frac{\sum_{n=0}^{N} \alpha_{n} \left( a-a_0\right)^n}{1+\sum_{m=1}^{M} \beta_{m}\left( a-a_0\right)^m} \ ,
\end{align}
where the truncation order is given by $N$ and $M$. The coefficients $\alpha_{n}$ and $\beta_m$ are assumed to be uniformly distributed in the range $[-1,1]$. We have tested that the results are not sensitive to changing the prior range. We also progressively raised the truncation order until the results converged, and adopted $N=M=9$. We consider, with equal weight, expansions around $a_0=0$ and $a_0=1$ to represent models that are close to LCDM in the past or at present, respectively. We also tried other parameterizations considered in \cite{Raveri:2017qvt}, such as polynomials in $(a-a_0)$, and found that the results are not sensitive to the choice.

To compute $\Sigma$ and $\mu$ and the expansion history, we use the publicly available EFTCAMB and EFTCosmoMC patches \cite{Hu:2013twa,Raveri:2014cka} to CAMB \cite{Lewis:1999bs} and CosmoMC \cite{Lewis:2002ah} (see~\cite{Hu:2014oga} for the implementation details). Given a choice of EFT functions, EFTCAMB first solves for the background evolution, then checks if conditions ensuring the stability of linear perturbations are satisfied, and then evolves such perturbations to evaluate the CMB spectra and other observables. Given the exact solutions for $\Delta$, $\Phi$ and $\Psi$ for a given model in the ensemble, we can calculate the exact $\mu(a,k)$ and $\Sigma(a,k)$ from Eqs.~(\ref{poisson_mg}) and (\ref{weyl_mg}) that define them. Alternatively, we can use EFTCAMB to perform the first two stages, \ie~to evolve the background and perform the stability check, and then evaluate $\Sigma$ and $\mu$ using the quasi-static (QS) expressions (\ref{eq:mu-horn}) and (\ref{eq:Sigma-horn}). For each sampling, we will present the results  for the exact and the QS $(\mu,\Sigma)$. By doing it both ways we can assess the validity of the QSA within Horndeski and also test the analytical arguments made in the previous Section under the QSA. 

In order for a model to be accepted by the sampler, it has to pass several checks. First, the model has to pass the stability conditions, as built in EFTCAMB. This filters out models with ghost and gradient instabilities in the scalar and tensor sectors. Further, we require viable models to fulfill weak observational and experimental priors on $\alpha_T(a)$, $\Omega(a)$ and $H(a)$. We emphasize that it is not our aim to perform a fit to data to derive observational bounds on $\Sigma$ and $\mu$. Instead, we want to derive theoretical priors on their values, but we want to exclude models that are in a gross violation of known constraints. The following priors simply require the realizations to be broadly acceptable:
\begin{itemize}
\item $\alpha_T(z=0) = 0$, to be consistent with the low redshift bounds on the speed of gravitational waves from GW170817 and GRB170817A \cite{TheLIGOScientific:2017qsa,Monitor:2017mdv,Coulter:2017wya};
\item $|\Omega(z=0)-1| < 0.1$, to be broadly consistent with the non-detection of the fifth force on Earth \cite{Brax:2012gr,Joyce:2014kja,Perenon:2015sla};
\item $|\Omega(z=1100)-1| < 0.1$, to comply with the BBN and CMB bounds constraining the value of the gravitational coupling to be within $10$\% of the Newton's constant measured on Earth \cite{Uzan:2010pm};
\item  $H(z)$ to be broadly consistent with existing cosmological distance measurements (see below for more details).
\end{itemize}
To dismiss expansion histories that are in gross disagreement with observations, we impose a weak Gaussian prior on $H(z)$ at several representative redshift values corresponding to existing luminosity distance measurements from supernovae and angular diameter distance measurements using Baryon Acoustic Oscillations (BAO). We take the prior to be peaked at $H(z)$ derived from the Planck 2015 best fit $\Lambda$CDM model \cite{Ade:2015xua}, with the standard deviation set at $30$\% of the peak value. The width of the prior is deliberately chosen to be wide enough to accommodate any tension existing between different datasets \cite{Zhao:2017cud}. The peak values of the $H(z)$ prior, along with the standard deviation, are plotted in Fig.~\ref{fig:H-prior}. We fix the spatial curvature to be zero, take the sum of neutrino masses to be $0.06$ eV, and impose conservative priors on the relevant cosmological parameters. Namely, the matter density fraction is allowed to change in the range $\Omega_{m}\in [0,1]$. Similarly, the present day dark energy fraction, which is not fixed by the flatness condition in non-minimally coupled models, was allowed to span $\Omega_{\rm DE}\in [0,1]$.

We then Monte Carlo sample the parameter space of all these models. To ensure a good coverage, we enforce a minimum number of $10^4$ accepted Monte Carlo samples. Depending on the acceptance rate, this results in $\sim 10^6 - 10^8$ of total samples. At each Monte Carlo step, after solving the background equations, we evaluate the stability of the corresponding model and, if this is found stable we compute the $\Sigma$ and $\mu$, sampling the $(a,k)$-plane at the following values:
\begin{eqnarray}
\nonumber
&&a \in \{ 0.25, 0.575, 0.9 \},\\
&&k \in \{ 0.001, 0.05, 0.1 \}, \nonumber
\end{eqnarray}
where $k$ has units of h/Mpc.

In order to study the effect of different EFT functions on the distribution of $\Sigma$ and $\mu$, we sample models from three different classes of theories. The first one is the class of generalized Brans-Dicke (GBD) which, in the EFT language, corresponds to having non-trivial functions $\Lambda$, $\Omega$ and $c$, while setting the rest to zero. The second is the $H_S$ class of models, with the unchanged speed of gravitational waves, which corresponds to adding non-trivial $M_2^4$ and $\bar{M}_1^3$ to the GBD functions. Finally, we consider the full class of Horndeski models, by adding a varying $\bar{M}_2^2$ to $H_S$, but we restrict  $\bar{M}_2^2$ to be zero at $z=0$, to comply with the strict bound on the gravitational wave speed today. The three classes of models are summarized in Table~\ref{tab:models}.

\subsection{Results of the numerical sampling}
 
\begin{figure}[htbp]
\includegraphics[width=0.45\textwidth]{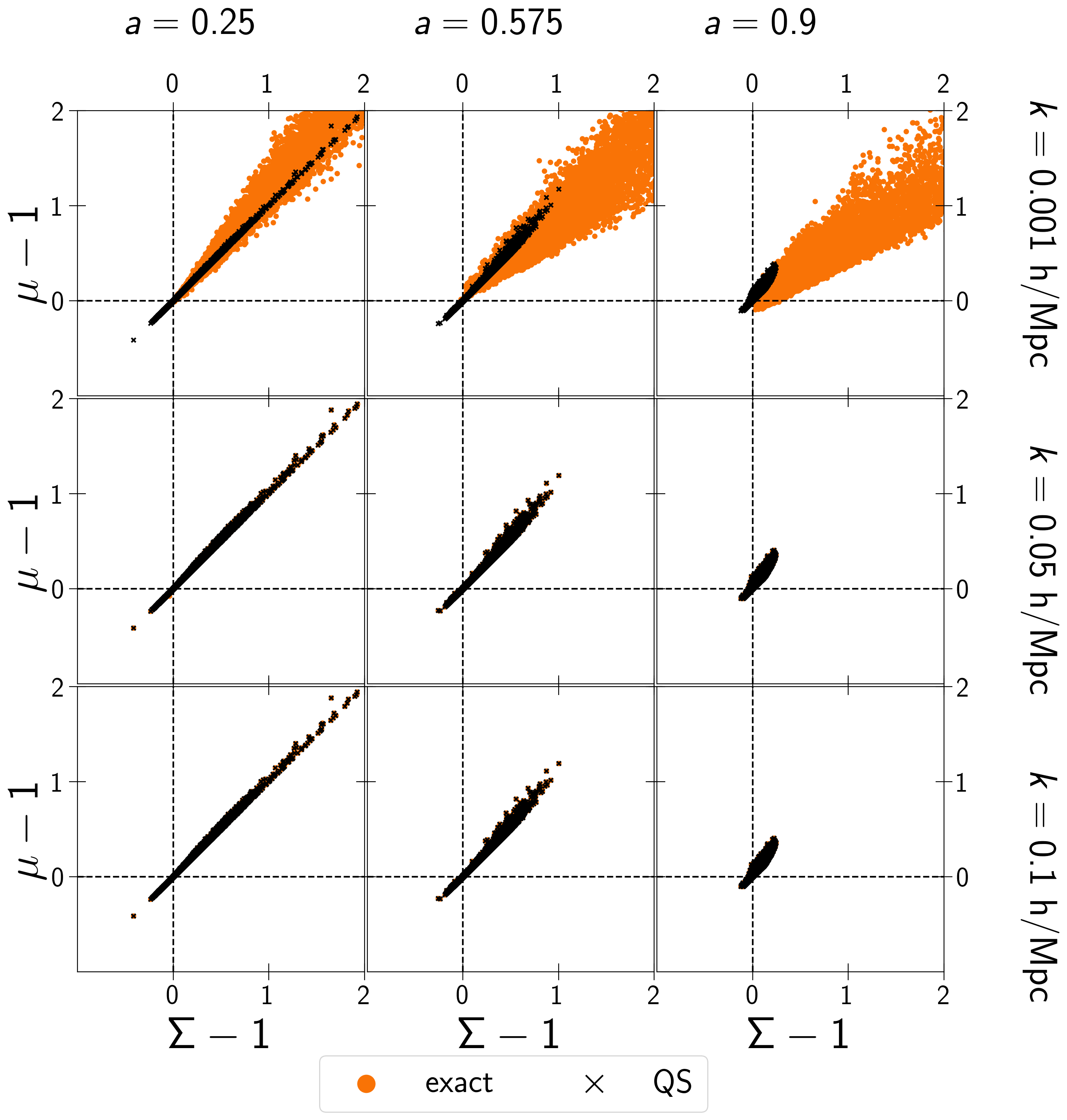}
\caption{Distributions of $\Sigma$ and $\mu$ in GBD models, {\it i.e.} the scalar-tensor models with a canonical kinetic term, at representative values of $a$ and $k$. Shown are results obtained by numerically solving exact equations for cosmological perturbations (orange dots) and by using the quasi-static (black crosses) forms of $\Sigma$ and $\mu$ given by Eqs.~(\ref{eq:mu-horn}) and (\ref{eq:Sigma-horn}).
\label{fig:gbd}}
\end{figure}

\begin{figure}[htbp]
\includegraphics[width=0.45\textwidth]{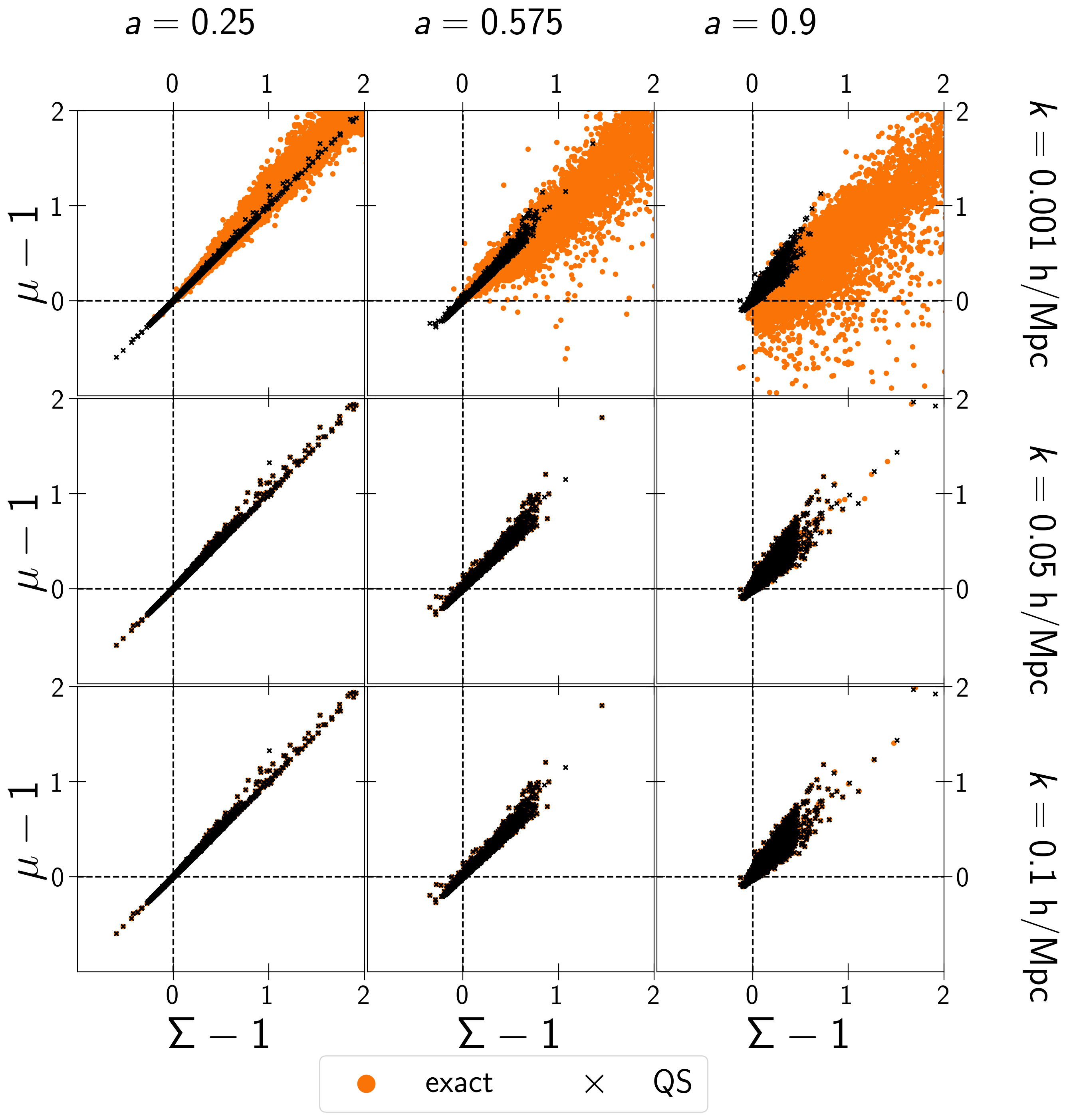}
\caption{Same as in Fig.~\ref{fig:gbd} but for the $H_S$ models, {\it i.e.} the subset of Horndeski models in which the speed of gravitational waves is the same as the speed of light at all redshifts.
\label{fig:hs}}
\end{figure}

\begin{figure}[htbp]
\includegraphics[width=0.45\textwidth]{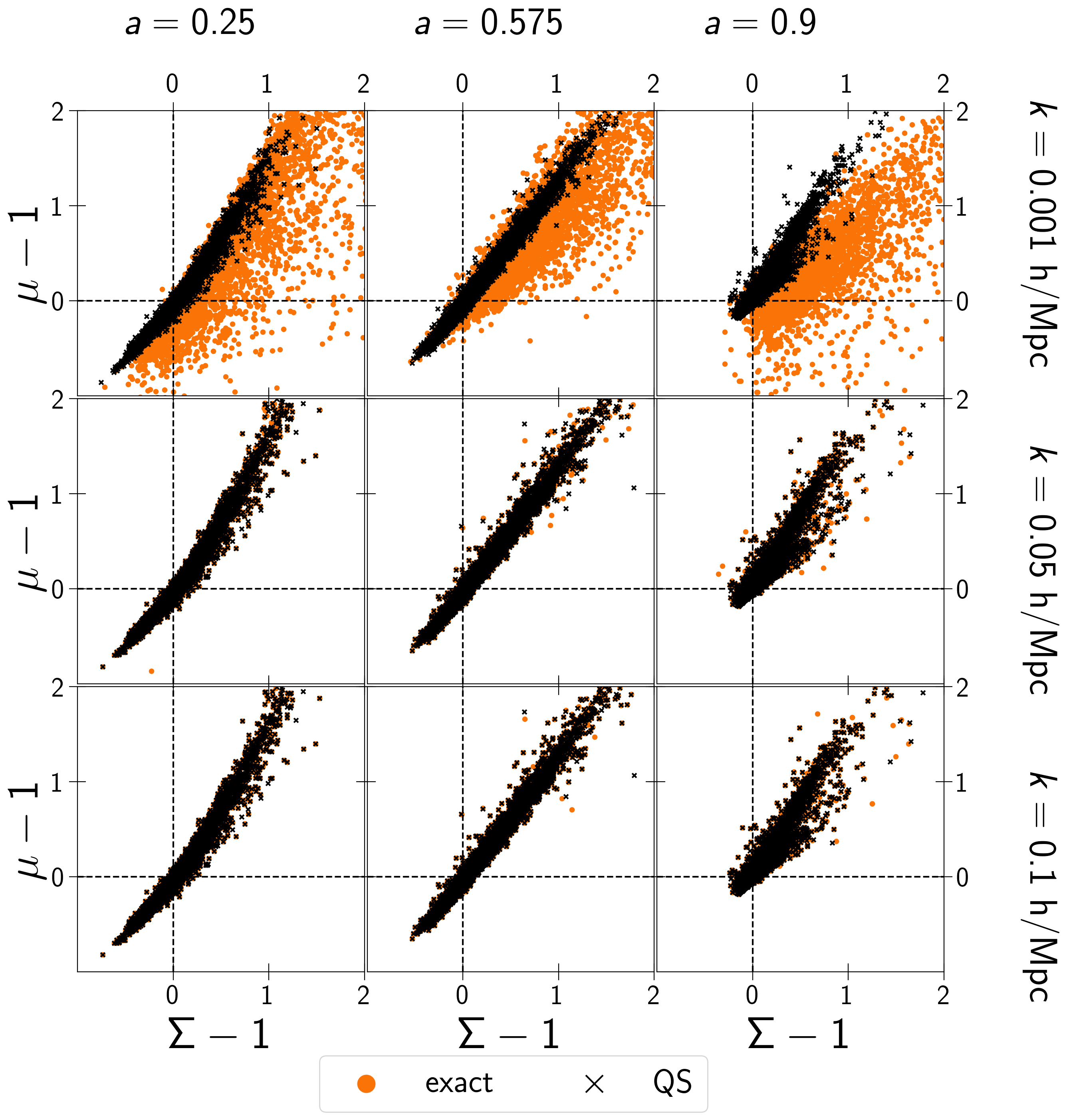}
\caption{Same as in Figs.~\ref{fig:gbd} and \ref{fig:hs}, but for the full class of Horndeski models with the restriction on variation of the speed of gravitational waves imposed only at $z=0$.
\label{fig:hor}}
\end{figure}

\begin{figure*}[htbp]
\includegraphics[width=0.9\textwidth]{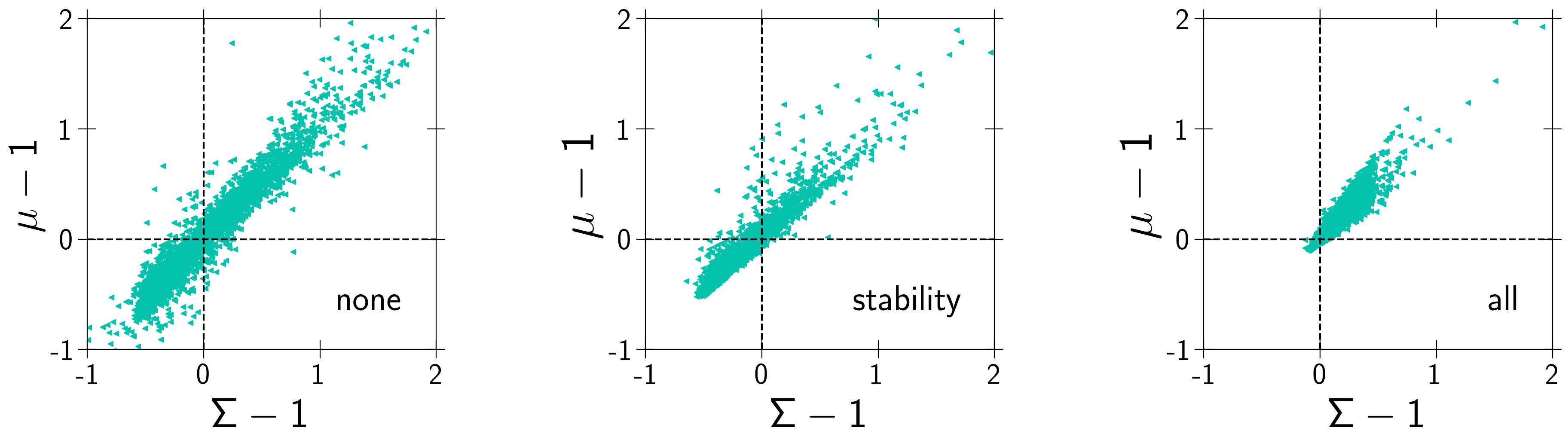}
\caption{Effects of imposing the stability conditions and observational priors on the $\Sigma$-$\mu$ distribution in the $H_S$ model for $a=0.9$ and $k=0.1$ h/Mpc. The three panels correspond to samples obtained in three different runs: sampling without any constraints (left panel), sampling with the stability constraints (middle), and sampling with both the stability constraints and observational priors (right). Each panel contains $10^4$ points. The impacts of stability and observational constraints shown here are representative of what happens at other redshifts and scales, and in the other classes of models that we studied. \label{fig:hs_stability}}
\end{figure*}

Figures~\ref{fig:gbd}, \ref{fig:hs} and \ref{fig:hor} show the numerically sampled distributions of $\Sigma$ and $\mu$ at representative values of $a$ and $k$  for GBD, $H_S$ and the full Horndeski model with the speed of gravity constrained to be unmodified today. In each figure, for the same ensemble of models, we show both the ``exact'' values, (calculated by numerically solving the full set of equations governing cosmological perturbations), as well as the values obtained using the QS expressions for $\Sigma$ and $\mu$ given by Eqs.~(\ref{eq:mu-horn}) and (\ref{eq:Sigma-horn}). We find that for all three models the QSA holds extremely well at $k=0.1$ and $0.05$ h/Mpc. Indeed, the clouds of exact and QS points effectively coincide for GBD and $H_S$, while for Horndeski there are only a few minor differences. We also see that, at $k=0.1, \ 0.05$ h/Mpc and at all redshifts, $\Sigma -1$ and $\mu -1$ are always of the same sign, following the conjecture made in \cite{Pogosian:2016pwr}.

The agreement between the exact and the QS calculations is much worse at $k=0.001$ h/Mpc, where we can see that the clouds of exact points are more spread compared to the QS clouds. A necessary condition for the QSA to hold is the requirement for the given Fourier mode to be inside the scalar field's sound horizon, \ie
\be
{k\over aH(a)} > c_s(a) \ ,
\label{QSA_necessary_condition}
\ee
where the speed of sound is given by Eq.~(\ref{eq:cs2}). In addition, the QSA assumes that the time-derivatives of the gravitational potentials and the scalar field perturbations are negligible compared to the spatial derivatives. To isolate the reason for the breakdown of the QSA at $k=0.001$ h/Mpc, we checked the fraction of models that pass the necessary condition (\ref{QSA_necessary_condition}) and found that only $1\%$ out of the total sample of $10^4$ models failed it.  This implies that for $k \lesssim 0.001$ h/Mpc one can no longer neglect the time-derivatives of the metric and field perturbations even on scales within the sound horizon of the scalar field. 

In the case of GBD, as seen in Fig.~\ref{fig:gbd}, the majority of both the QS and the exact values satisfy $(\Sigma -1)(\mu -1) \ge 0$. Only about $1\%$ of exact points in the $k=0.001$ h/Mpc, $a=0.9$ panel violate the conjecture, with no violations seen in the other panels. For $H_S$, the conjecture holds very well for the QS points, but not always for the exact points. We find that about $10\%$ of the exactly calculated points fall in the bottom-right quadrant at late redshifts and large scales, i.e. in the $k=0.001$ h/Mpc, $a=0.9$ panel, with only a handful of points violating the conjecture at higher redshifts for $k=0.001$ h/Mpc. Finally, for the full Horndeski sampling, we again find that the conjecture holds well under the QSA, and for the exact points on smaller scales ($k=0.1$ and $0.05$ h/Mpc). However, about $10\%$ of the models violate the conjecture at all three values of $a$ for $k=0.001$ h/Mpc. It is interesting to notice that, in those cases, the conjecture is always violated in the same direction, with a positive $\Sigma -1$ and a negative $\mu -1$.

In Fig.~\ref{fig:hs_stability}, we show the effects of imposing the stability constraints and observational priors on the distribution of $\Sigma$ and $\mu$. We consider the case of the $H_S$ model at $k=0.1$ h/Mpc and $a=0.9$, which is representative of the trends we see at other scales and redshifts and in the other models. The three panels show samples of the $H_S$ models without imposing any constraints (left panel), after filtering out models with the ghost and gradient instabilities \cite{Hu:2014oga}  (middle panel) and after imposing both the stability constraints and observational priors (right panel). In each case, we run the simulation until $10^4$ ``successful'' models are accumulated. From these plots, we can see that imposing the stability conditions removes all points from the bottom-right quadrant. As discussed in Section~\ref{sec:conjectureC2}, this happens because stability requires $c^2_s \alpha >0$. Finally, in the right panel, we see that adding the observational priors eliminates the models belonging to the top-left quadrant. This confirms the argument made in Section~\ref{sec:conjectureC1} according to which getting $\Sigma < 1 $ and $\mu > 1$ would require large variations in $\Omega$, which are indeed strongly suppressed by the observational constraints defined in the beginning of this Section. We note that the points in the middle and the right panels are not simple subsets of the left panel, since we run the simulation until the same number of points is accumulated in each case.

From Fig.~\ref{fig:hs_stability} we also notice that the combined effect of the stability conditions and the observational priors is to drastically reduce the models in the bottom-left quadrant, where $\mu-1<0$ and $\Sigma-1<0$. In the absence of ghosts, the scalar force is always attractive, thus the fifth force contribution generally favours $\mu >1$. One could still have $\mu<1$, driven by the $1/\Omega$ factor in the QS expression (\ref{mu_infty_hs}) for $\mu$, \ie~ having $\Omega$ that is significantly greater than $1$ can result in $\mu<1$. However, observational constraints restrict $\Omega \sim 1$ at late times, making it difficult to get $\mu<1$. We see in Fig.~\ref{fig:hs} that the bottom-left quadrant has practically no points at $a=0.9$, but is more populated at earlier times, since the observational constraint on $\Omega$ are weaker at higher redshifts.

\section{Summary and Conclusions}
\label{sec:summary}

We studied the range of values that phenomenological functions $\Sigma$ and $\mu$ can take in viable Horndeski theories. To do so, we built numerical samples of Horndeski models that pass the no ghost and no gradient instability constraints, as well as a set of weak observational constraints. For each model, we computed $\Sigma$ and $\mu$ by numerically solving the exact equations for cosmological perturbations, and also by using the analytical expressions of $\Sigma$ and $\mu$ derived under the QSA. This allowed us to check the validity range of the quasi-static approximation (QSA), as well as the validity of the conjecture made in~\cite{Pogosian:2016pwr} that $(\Sigma-1)(\mu -1) \ge 0$ in viable Horndeski theories.

We find that the QSA holds really well at small and intermediate scales, but breaks down at $k\lesssim 0.001$ h/Mpc. This happens despite the fact that the Fourier modes in question are still well-within the scalar field's sound horizon. Instead, it is due to the time-derivatives of the metric and the scalar field perturbations, which are neglected under the QSA, becoming comparable to the spatial derivatives.

We have considered three types of Horndeski theories summarized in Table~\ref{tab:models}: the Generalized Brans-Dicke (GBD) models, \ie~models with a canonical form of the scalar field kinetic energy term, the $H_S$ class of models, with the unchanged speed of gravitational waves, and the full class of Horndeski models with the speed of gravity constrained to be the same as the speed of light at present epoch, to comply with the strict bound on the gravitational wave speed at $z<0.01$ from GW170817 and GRB170817A \cite{TheLIGOScientific:2017qsa,Monitor:2017mdv,Coulter:2017wya}.

We find that the $(\Sigma-1)(\mu -1) \ge 0$ conjecture holds very well for the GBD models. It also holds very well for the other two classes of models within the QSA, but the exact calculations show that about $10\%$ of $H_S$ and Horndeski models violate the conjecture at $k=0.001$ h/Mpc, with $\Sigma >1$ and $\mu <1$. 

We analytically examined the conditions under which $(\Sigma-1)(\mu -1) \ge 0$ can be violated, separately considering the QS expressions for $\Sigma$ and $\mu$ on the super-Compton and sub-Compton limits. We identified the important role played by the no ghost and no gradient instability conditions in preventing values in the $\Sigma >1$ and $\mu <1$ range. We have also highlighted the importance of the constraints on the variation of the gravitational coupling in ensuring the $(\Sigma-1)(\mu -1) \ge 0$ trend. Since the variation of the gravitational coupling affects the background expansion history, constraints on the latter contribute to restricting the range of $\Sigma$ and $\mu$ values. This effect was not included in an earlier study of correlations between $\Sigma$ and $\mu$ \cite{Perenon:2016blf} that was based on a framework in which the expansion history was assumed to be known independently from the functions controlling the evolution of perturbations.
Our analysis shows that, when searching for signatures of MG, the expansion history should be co-varied with $\Sigma$ and $\mu$ aided by weak theoretical priors based on broad classes of theories. Studies like this, and the one in \cite{Raveri:2017qvt}, could be used to build such theoretical priors.

Our study demonstrates the benefits and the complementarity of different frameworks for testing scalar-tensor alternatives to GR. Phenomenological functions such as $\Sigma$ and $\mu$ are closely related to observations and can be directly fit to data using simple parameterizations. However, there is no guarantee that their best fit values would be consistent with theory. On the other hand, fitting the EFT functions of (\ref{EFT_action}) or the Unified functions of (\ref{alpha_action}) directly to data is not practical, as there are many degeneracies and the outcome strongly depends on the assumed functional form. Instead, the EFT framework can be used to systematically generate viable Horndeski theories and derive theoretical priors on $\Sigma$ and $\mu$, similarly to how it was done in this study. The Unified framework is highly complementary, allowing to derive simple QS forms of $\Sigma$ and $\mu$ that make it easier to interpret the numerical results analytically. 

This work shows that, even with the strict bound on the present day gravitational wave speed, there is still room within Horndeski theories for non-trivial signatures of modified gravity that can be measured at the level of linear perturbations. Moreover, there are clear correlations between the phenomenological functions $\Sigma$ and $\mu$ that can help to determine if a potentially measured departure from LCDM is consistent with a scalar-tensor theory.

\acknowledgments
We thank Louis Perenon and Federico Piazza for useful discussions. 
SP and AS acknowledge support from the NWO and the Dutch Ministry of Education, Culture and Science (OCW), and also from the D-ITP consortium, a program of the NWO that is funded by the OCW. KK is supported by the STFC grant ST/N000668/1. The work of KK has also received funding from the European Research Council (ERC) under the European Union's Horizon 2020 research and innovation programme (grant agreement 646702 ``CosTesGrav").
The work of LP is supported by the Natural Sciences and Engineering Research Council of Canada (NSERC). MR is supported by U.S. Dept. of Energy contract DE-FG02-13ER41958. 
\appendix

\section{Relevant Equations}
\label{app:hor}

Under the QSA, the equations of motion for perturbations in Horndeski theories can be written as \cite{Bloomfield:2012ff}
\ba
&&A_1 {k^2 \over a^2} \Phi + A_2 {k^2 \over a^2} \pi = -\rho \Delta ,
\label{eom_poisson}
\\
&&B_1 \Psi + \Phi + B_3 \pi = 0,
\label{eom_anisotropy}
\\
&&C_1 {k^2 \over a^2} \Phi + C_2 {k^2 \over a^2} \Psi + \left(C_3 {k^2 \over a^2} + C_\pi  \right) \pi = 0,
\label{eom_pi}
\ea
where 
\ba
A_1 &=& 2(m_0^2\Omega+{\bar M}^2_2) \nonumber \\
A_2 &=& -m_0^2 \dot{\Omega}-\bar{M}^3_1 \nonumber \\
B_1 &=& -{ m_0^2\Omega +{\bar M}^2_2 \over m_0^2\Omega} \nonumber \\
B_3 &=& -{m_0^2 \dot{\Omega}  + (H + \partial_t){\bar M}^2_2  \over m_0^2 \Omega} \nonumber \\
C_1 &=& m_0^2 \dot{\Omega}  + (H + \partial_t){\bar M}^2_2 \nonumber \\
C_2 &=& - {1 \over 2}(m_0^2 \dot{\Omega}+\bar{M}^3_1)  \nonumber \\
C_3 &=& c-{1\over 2} (H+\partial_t)\bar{M}_1^3 + (H^2+\dot{H}+H\partial_t){\bar M}^2_2 \nonumber \\
C_\pi &=& {m_0^2\over 4}\dot{\Omega}\dot{R}^{(0)}-3c\dot{H} + {3\over 2} ( 3H\dot{H} +\dot{H} \partial_t+\ddot{H}) \bar{M}_1^3 
\nonumber \\
&+& 3 \dot{H}^2 \bar{M}_2^2 
\label{M2_Hornd}
\ea
The phenomenological functions $\mu$ and $\Sigma$ can be written as
\ba
4\pi G \mu = {\mu \over 2m_0^2}&=&{f_1+f_2 \ a^2/k^2 \over f_3+f_4 \ a^2/k^2},
\label{eq:mu-eft} \\
8\pi G \Sigma={\Sigma \over m_0^2} &=& {f_1+f_5+(f_2+f_6) a^2/k^2 \over f_3+f_4 \ a^2/k^2},
\label{eq:Sigma-eft}
\ea
where
\ba
f_1&=&C_3-C_1B_3 \nonumber \\
f_2&=&C_\pi \nonumber \\
f_3&=&A_1(B_3C_2-B_1C_3)+A_2(B_1C_1-C_2) \nonumber \\
f_4&=& -A_1B_1C_\pi \nonumber \\
f_5&=&  B_3C_2-B_1C_3 \nonumber \\
f_6&=&  -B_1 C_\pi
\label{eq:f_i}
\ea
The functions appearing in the ``Unified'' action (\ref{alpha_action}) are related to the functions appearing in the EFT action (\ref{EFT_action}) via \cite{Bellini:2014fua}
\ba
M_*^2 &=& m_0^2\Omega + \bar{M}^2_2 
\label{eq:M*} \\ 
HM_*^2 \alpha_M &=& m_0^2\dot{\Omega} + \dot{\bar{M}}^2_2 \\
M_*^2 \alpha_T &=&  -\bar{M}^2_2 
\label{eq:alphaT} \\
HM_*^2 \alpha_B &=& -m_0^2\dot{\Omega} - \bar{M}^3_1 
\label{eq:alphaB} \\
H^2M_*^2 \alpha_K &=& 2c+4M_2^4 \label{eq:alphaK} \\\ .
\ea
These are related to the functions in the original Horndeski Lagrangian (\ref{Horn_Li}) via \cite{Bellini:2014fua}
\ba
&&M_*^2 = 2[G_4-2XG_{4X}+XG_{5\phi}-{\dot \phi}HXG_{5X}] \\
&&HM_*^2 \alpha_M = {d M_*^2 \over dt} \\
&&M_*^2 \alpha_T = 2X[2G_{4X}-2G_{5\phi}-(\ddot{\phi} - H\dot{\phi}) G_{5X}] \\ \nonumber
&&HM_*^2 \alpha_B = 2\dot{\phi} [XG_{3X}-G_{4\phi}-2XG_{4\phi X}] \\ \nonumber
&& \ \ \ +8XH(G_{4X}+2XG_{4XX}-G_{5\phi}-XG_{5\phi X}) \\
&& \ \ \ + 2\dot{\phi}XH^2[3G_{5X}+2XG_{5XX}] \\ \nonumber
&&HM_*^2 \alpha_K = 2X[K_X+2XK_{XX}-2G_{3\phi}-2XG_{3\phi X}] \\
&& \ \ \ + 12\dot{\phi}XH[G_{3X}+XG_{3XX}-3G_{4\phi X}-2XG_{4\phi XX}] \nonumber \\
&& \ \ \ +12XH^2[G_{4X}+8XG_{4XX} + 4X^2 G_{4XXX}] \nonumber \\
&& \ \ \ -12XH^2[G_{5\phi} +5XG_{5\phi X}+2X^2G_{5\phi XX}] \nonumber \\
&& \ \ \ +4\dot{\phi}XH^3[3G_{5X}+7XG_{5XX}+2X^2G_{5XXX}]
\ea
The speed of sound of the scalar field perturbations is given by
\ba
\nonumber
c_s^2&=&{2\over \alpha} \Big[ \left(1-{\alpha_B \over 2} \right) \Big( \alpha_M-\alpha_T+ {\alpha_B \over 2 }(1+\alpha_T) -{\dot{H} \over H^2} \Big) \\
&+&{\dot{\alpha}_B \over 2H} - {\rho_m+P_m \over 2M^2_*H^2}  \Big] \ ,
\label{eq:cs2}
\ea
where $\alpha = \alpha_K+3 \alpha_B^2/2$.

\bibliography{sigmamu}

\end{document}